\documentclass[11pt]{article}
\usepackage{amsfonts,colordvi}
\usepackage{amssymb}
\newcommand{\eq}{\begin{equation}}
\newcommand{\en}{\end{equation}}
\newcommand{\eqn}{\begin{eqnarray}}
\newcommand{\enn}{\end{eqnarray}}

\newcommand{\beq}{\begin{equation}}
\newcommand{\eeq}{\end{equation}}
\newcommand{\M}{\ensuremath{\mathcal{M}}}
\newcommand{\CN}{\ensuremath{\mathcal{N}}}
\newcommand{\tn}{\ensuremath{\tilde{n}}}
\newcommand{\ta}{\ensuremath{\tilde{a}}}
\newcommand{\tb}{\ensuremath{\tilde{b}}}

\newcommand{\tx}{\ensuremath{\tilde{x}}}
\newcommand{\ty}{\ensuremath{\tilde{y}}}
\newcommand{\ti}{\ensuremath{\tilde{I}}}
\newcommand{\tj}{\ensuremath{\tilde{J}}}
\newcommand{\tk}{\ensuremath{\tilde{K}}}
\newcommand{\tl}{\ensuremath{\tilde{L}}}

\begin{document}

\begin{titlepage}

\begin{center}

\hfill hep-th/0304109\\
\hfill CERN-TH/2003-085
\vskip 1cm
\begin{LARGE}
\textbf{Unified 
Maxwell-Einstein and\\ \vspace{1mm}
Yang-Mills-Einstein Supergravity
Theories\\\vspace{2mm} in  Five Dimensions   } \footnote{ Work supported
  in part by the National
Science Foundation under Grant Number PHY-0099548.}
\end{LARGE}\\
\vspace{1.0cm}
\begin{large}
M. G\"{u}naydin$^{\dagger}$ \footnote{murat@phys.psu.edu} and
M. Zagermann$^{\ddagger}$ \end{large}\footnote{Marco.Zagermann@cern.ch}  \\
\vspace{.35cm}
$^{\dagger}$ \emph{Physics Department \\
Pennsylvania State University\\
University Park, PA 16802, USA} \\
\vspace{.3cm}
and \\
\vspace{.3cm} $^{\ddagger}$ \emph{CERN, Theory Division \\
CH-1211, Geneva 23, Switzerland } \\
\vspace{0.5cm} {\bf Abstract}
\end{center}
\begin{small}
Unified $\CN=2$   Maxwell-Einstein supergravity theories (MESGTs)
are supergravity theories in which all the vector fields, including
the graviphoton,
transform in an irreducible representation of a simple global
symmetry group of the
 Lagrangian.
As was established long time ago, in five dimensions there exist only four
unified Maxwell-Einstein supergravity theories
whose target manifolds are symmetric spaces.
These theories are defined by
the four simple Euclidean Jordan algebras of degree three.
In this paper, we show that, in addition to these four
unified MESGTs with symmetric target spaces, there exist
three infinite families of unified MESGTs as well as
 another exceptional one. These novel unified  MESGTs are
 defined by non-compact (Minkowskian) Jordan algebras, and
 their target spaces
are in general neither symmetric nor homogeneous. The members of
one of these three infinite
families can be gauged in such a way as to obtain an infinite family of
{\it unified} $\CN=2$ Yang-Mills-Einstein supergravity theories, in which all
vector fields transform in the adjoint representation of a simple
 gauge group of the
type $SU(N,1)$.
The corresponding gaugings in the other two infinite families lead to
Yang-Mills-Einstein supergravity theories coupled to tensor
multiplets.

\end{small}

\end{titlepage}

\renewcommand{\theequation}{\arabic{section}.\arabic{equation}}
\section{Introduction}
\setcounter{equation}{0}

One of the original   motivations for studying  supersymmetric
theories in  particle physics was the hope that they might provide
the right framework for \\
a) unifying gravity with the gauge interactions of the Standard Model and\\
b) for putting the fermionic matter constituents on an equal
footing with the bosonic fields that mediate the interactions
between them.

In the early 1980s,  efforts in this direction
  culminated  in  the construction of
the maximally supersymmetric, $\mathcal{N}=8$ supergravity theory
with the gauge group $SO(8)$ \cite{dWN}. In this theory, all
fields sit in one and the same supermultiplet and are thus
connected by supersymmetry and/or gauge transformations.

It soon became clear, however, that neither the $\mathcal{N}=8$
theory nor any other extended  four-dimensional (4D)
  supergravity theory can be a
phenomenologically realistic model of low energy particle physics,
and that, instead, a
realistic four-dimensional extension of the Standard Model can
involve at most minimal $\mathcal{N}=1$ supersymmetry. Moreover,
gravity, the Yang-Mills gauge fields and the matter constituents  all
have to sit in different types of supermultiplets and can
therefore not be connected by supersymmetry transformations or any
other obvious symmetry. Thus, the original idea of using
supersymmetry to directly unify all particles and interactions in
terms of a purely 4D field theory did not prove to be successful.

Nevertheless, supersymmetry still plays an important r\^{o}le  in the 
context of
unification, albeit now  in   a more
indirect way. For one thing, supersymmetry naturally appears in string 
theories,
i.e., in  the most promising
known models for a complete unification of all  particles and interactions,
including gravity. Furthermore, from
a more bottom-up point of view, a supersymmetrization of the Standard Model
spectrum seems to be required in
order  to reconcile precision measurements at particle colliders with  the
idea of converging Standard Model
couplings within a conventional GUT  scenario \cite{gutscenario}.

Partly motivated by certain string theory constructions, such grand unified
models have recently also been studied
within a higher dimensional framework in order to overcome some of
the notorious problems of   standard 4D
GUTs, such as proton decay or the doublet-triplet splitting problem.
Especially five-dimensional models have been studied
quite extensively  in this context starting with refs. \cite{fivedguts}.

In five dimensions, the smallest possible amount of supersymmetry
involves eight real supercharges, which, in analogy with the
corresponding 4D terminology, is   often referred to as
$\mathcal{N}=2$ supersymmetry.
Unlike its minimally (i.e., $\mathcal{N}=1$) supersymmetric
 counterpart in 4D, the
5D, $\mathcal{N}=2$ supergravity multiplet contains a vector field
(the `graviphoton'). It is therefore, in principle,
 conceivable to have an
additional bosonic symmetry that could map the 5D graviphoton to some
or all of the vector fields that sit in 5D vector multiplets. 
As supersymmetry interpolates between the graviphoton and the graviton, 
one might then, in a certain sense,   view such a symmetry
as a `unification' of the vector multiplet sector and the gravity
sector of the theory.

Such extra bosonic symmetries are of course  nothing new, but constitute
a  well-known feature
of extended
supergravity theories, already
  in four dimensions. Consider for example
 4D,
$\mathcal{N}=4$ supergravity coupled to $n$ Abelian vector
multiplets \cite{n4d4sugra}. 
This theory contains $(6+n)$ vector fields, where 6
come from the supergravity multiplet (i.e., there are
 six `graviphotons') and the remaining $n$ are supplemented by the
$n$ vector multiplets. In addition to the local $\mathcal{N}=4$ supersymmetry,
this theory has a global symmetry
group of the form $G=SU(1,1)\times SO(6,n)$. 
Under the $SO(6,n)$
factor, the $(6+n)$ vector
fields of the theory transform irreducibly in the $\mathbf{(6+n)}$ 
representation, i.e.,
they are all connected
by a symmetry of the theory even though they originate from different 
types of  supermultiplets.
In the following, we will
call such an extended Abelian supergravity theory in which  \emph{all} 
the vector fields transform
\emph{irreducibly} under a \emph{simple} global symmetry group `unified', 
or, more
precisely, a `unified'
Maxwell-Einstein supergravity theory (unified MESGT). As the above example
illustrates, the `unifying' symmetry
group of a unified \ MESGT is, in general, non-compact.

One might wonder whether it is also possible to have a similar
`unification' between  graviphotons and vector fields from vector
multiplets, when the latter gauge a non-Abelian Yang-Mills
symmetry. Let us first reconsider the above $\mathcal{N}=4$
theories. For the special case $n=3$, the global $SO(6,3)$ symmetry
has the obvious subgroup $SO(2,1)\times SO(2,1)\times SO(2,1)$,
under which the original $\textbf{(6+3)}$ of $SO(6,3)$ decomposes
into three $SO(2,1)$ triplets. Using standard supergravity
techniques, one can then turn $SO(2,1)\times SO(2,1)\times
SO(2,1)$ into a Yang-Mills-type gauge symmetry under which all
vector fields transform irreducibly in the adjoint representation.
However, this
gauge group is not \emph{simple}, so that, in analogy to our Abelian
definition, we shall not call this theory `unified'. Instead, we
define a unified Yang-Mills-Einstein supergravity theory  (unified YMESGT)
to be a supergravity theory in which a \emph{simple}
Yang-Mills-type gauge group acts \emph{irreducibly} on all
graviphotons and all the vector fields that come from vector
multiplets.

It is  easy to convince oneself that such a theory cannot be constructed 
for our above 4D,  $\mathcal{N}=4$
examples, no matter how many vector multiplets are used. One might 
therefore wonder whether such unified YMESGTs
exist at all.

 As was first shown in  ref. \cite{GST4}, the answer is in the affirmative.
The example constructed in \cite{GST4} (see also \cite{GST2})
  describes the coupling of 5D, 
$\mathcal{N}=2$ supergravity to 14   vector
multiplets and has the gauge group $SU(3,1)$. This gauge group is 
possible because the graviphoton of the
$\mathcal{N}=2$ supergravity multiplet and the 14 vector fields of 
the 14  vector multiplets combine into the
15-dimensional adjoint representation of $SU(3,1)$. Turning off the 
Yang-Mills coupling, one obtains a unified
MESGT in which the global unifying group gets enhanced to $SU^{*}(6)$. 
In \cite{GST1}, this unified MESGT with 14
vector multiplets was found  to be   a member of a family of four unified 
MESGTs, which have, respectively,  5,
8, 14 or 26 vector multiplets and are associated with the  simple Jordan 
algebras of Hermitian $(3\times
3)$-matrices over the four division algebras $\mathbb{R}$, $\mathbb{C}$, 
$\mathbb{H}$ and $\mathbb{O}$. If one
restricts oneself to theories in which the scalar manifold forms a 
symmetric space, these four theories are the
only unified MESGTs in 5D. Except for the theory with 14 vector multiplets, 
none of them can be turned into a
unified YMESGT upon gauging a subgroup of the relevant global symmetry 
groups. In the class of theories with
symmetric target spaces, this theory is thus unique.

In this paper we   show that if one abandons the restriction to 
\emph{symmetric} target spaces, there are three
infinite families of novel unified MESGTs in 5D, $\mathcal{N}=2$ 
supergravity as well as one novel exceptional
unified MESGT. One of the three infinite families can even be turned 
into an infinite family of unified YMESGTs
with gauge groups of the type $SU(1,N)$ for arbitrary high $N\geq 2$. 
We can show that these theories exhaust all
possible unified YMESGTs in five dimensions (the theory found in 
\cite{GST4,GST2} turns out to be a special case of
this infinite family). As a by-product, we find an intriguing connection 
to a classical work by Elie Cartan on
some remarkable families of isoparametric hypersurfaces in spaces of 
constant curvature \cite{cartan} .

The paper is organized as follows. In Section 2, we briefly recall 
the basic properties of $\mathcal{N}=2$ MESGTs
in five dimensions. Section 3 gives a short description of the four 
unified MESGTs with symmetric target spaces
that were found in
 ref. \cite{GST1}.
In Section 4 we recall the relation of these four theories to
Jordan algebras and show
how the language of Jordan algebras quickly leads to the
construction of three novel
infinite families as well as another exceptional unified MESGT.
The members of only one of the three infinite families
can be turned into unified  YMESGTs, which is further explained in Section 5.
In the course of this work, we became aware of a connection
to earlier work by E. Cartan on isoparametric hypersurfaces in spaces of 
constant
curvature \cite{cartan}. This connection is sketched in Section 6. Our results
are summarized and discussed in Section 7.



\section{5D, $\mathcal{N}=2$ Maxwell-Einstein supergravity theories}
\setcounter{equation}{0}

In this section, we review the salient
features of general  5D,  $\mathcal{N}=2$ Maxwell-Einstein
supergravity theories (MESGTs)
\cite{GST1}.\footnote{ Our conventions coincide with those of
ref. \cite{GST1,GST2,GZ1}. In particular, we will use the
mostly positive metric signature $(-++++)$ and impose the
`symplectic' Majorana condition on all fermionic quantities.}

A 5D, $\mathcal{N}=2$ MESGT describes
the coupling of pure 5D, $\mathcal{N}=2$
 supergravity to an arbitrary number, $\tn$, of vector
multiplets.
The fields of the  supergravity multiplet  are the
f\"{u}nfbein $e_{\mu}^{m}$, two gravitini $\Psi_{\mu}^{i}$
($i=1,2$) and one vector field $A_{\mu}$ (the graviphoton).
An $\mathcal{N} =2$ vector multiplet contains a
vector field $A_{\mu}$, two spin-$1/2$ fermions $\lambda^{i}$ and
one real scalar field $\varphi$. The fermions of each of these
multiplets transform as doublets under the $USp(2)_{R}\cong
SU(2)_{R}$ R-symmetry group of the $\mathcal{N} =2$ Poincar\'{e}
superalgebra;  all other fields are $SU(2)_{R}$-inert.

Putting everything together, the total field content
of an $\mathcal{N}=2$ MESGT is thus
\begin{equation}
\{ e_{\mu}^{m}, \Psi_{\mu}^{i}, A_{\mu}^{\ti}, \lambda^{i\ta}, \varphi^{\tx}\}
\end{equation}
with
\begin{eqnarray*}
\ti&=& 0,1,\ldots, \tn\\
\ta&=& 1,\ldots, \tn\\
\tx&=& 1,\ldots, \tn.
\end{eqnarray*}
Here,
 we have combined the graviphoton with the $\tn$ vector fields of the $\tn$
vector multiplets into a single $(\tn+1)$-plet of vector fields
$A_{\mu}^{\ti}$ labelled
by the index $\ti$. The indices $\ta, \tb, \ldots$ and $\tx, \ty,
\ldots$ denote the
flat and curved indices, respectively, of the
 $\tn$-dimensional target manifold, $\mathcal{M}$,
of the scalar fields.

The bosonic part of the Lagrangian is
given by (for the fermionic part and further details see \cite{GST1})
\begin{eqnarray}\label{Lagrange}
e^{-1}\mathcal{L}_{\rm bosonic}&=& -\frac{1}{2}R
-\frac{1}{4}{\stackrel{\circ}{a}}_{\ti\tj}F_{\mu\nu}^{\ti}
F^{\tj\mu\nu}-\frac{1}{2}g_{\tx\ty}(\partial_{\mu}\varphi^{\tx})
(\partial^{\mu}
\varphi^{\ty})+\nonumber \\ &&+
 \frac{e^{-1}}{6\sqrt{6}}C_{\ti\tj\tk}\varepsilon^{\mu\nu\rho\sigma\lambda}
 F_{\mu\nu}^{\ti}F_{\rho\sigma}^{\tj}A_{\lambda}^{\tk},
\end{eqnarray}
where  $e$ and $R$ denote the f\"{u}nfbein determinant
and the scalar curvature, respectively, and
$F_{\mu\nu}^{\ti}$ are the Abelian field strengths of the vector
fields $A_{\mu}^{\ti}$.
The metric, $g_{\tx\ty}$, of the scalar manifold $\M$
 and the matrix ${\stackrel{\circ}{a}}_{\ti\tj}$ both depend
on the scalar fields $\varphi^{\tx}$. The completely symmetric
tensor $C_{\ti\tj\tk}$, by contrast, is constant.
Remarkably,  the entire $\mathcal{N}=2$
MESGT (including also the fermionic terms and the
supersymmetry transformation laws we have not shown here) is uniquely
determined by the   $C_{\ti\tj\tk}$ \cite{GST1}. More explicitly, the
$C_{\ti\tj\tk}$ define a cubic polynomial, $\mathcal{V}(h)$,
 in $(\tn+1)$
real variables $h^{\ti}$ $(\ti=0,1,\ldots,\tn)$,
\begin{equation}
\mathcal{V}(h):=C_{\ti\tj\tk}h^{\ti}h^{\tj}h^{\tk}\ .
\end{equation}
This polynomial defines a metric, $a_{\ti\tj}$,
 in the (auxiliary) space $\mathbb{R}^{(\tn+1)}$ spanned by the $h^{\ti}$:
\begin{equation}\label{aij}
a_{\ti\tj}(h):=-\frac{1}{3}\frac{\partial}{\partial h^{\ti}}
\frac{\partial}{\partial h^{\tj}} \ln \mathcal{V}(h)\ .
\end{equation}
The  $\tn$-dimensional    target space, $\mathcal{M}$, of the scalar
fields $\varphi^{\tx}$ can then be represented as the hypersurface
\cite{GST1}
\begin{equation}\label{hyper}
{\cal V} (h)=C_{\ti\tj\tk}h^{\ti}h^{\tj}h^{\tk}=1 \ ,
\end{equation}
with $g_{\tx\ty}$ being the pull-back of (\ref{aij}) to $\mathcal{M}$.
The quantity ${\stackrel{\circ}{a}}_{\ti\tj}(\varphi)$ appearing in
(\ref{Lagrange}), finally, is given by the componentwise restriction of
$a_{\ti\tj}$ to $\mathcal{M}$:
\[
{\stackrel{\circ}{a}}_{\ti\tj}(\varphi)=a_{\ti\tj}|_{{\cal V}=1} \ .
\]

The physical requirement of unitarity requires
 $g_{\tx\ty}$ and ${\stackrel{\circ}{a}}_{\ti\tj}$ to  be positive definite.
This requirement induces constraints on the possible $C_{\ti\tj\tk}$,
and in \cite{GST1} it   was shown that any $C_{\ti\tj\tk}$ that
satisfy these constraints can be brought to the following form
\begin{equation}\label{canbasis}
C_{000}=1,\quad C_{0ij}=-\frac{1}{2}\delta_{ij},\quad  C_{00i}=0,
\end{equation}
with  the remaining coefficients $C_{ijk}$
 ($i,j,k=1,2,\ldots , \tn$) being  completely arbitrary.
 We shall refer to this basis as the canonical basis.
The arbitrariness of the  $C_{ijk}$ in the canonical basis
implies  that,  for a fixed
number $\tn$ of vector multiplets, different
target manifolds $\mathcal{M}$  are, in
general,  possible.


\section{Unified MESGTs
with symmetric target spaces}
\setcounter{equation}{0}

In this paper, we are interested in `unified'
 Maxwell-Einstein supergravity
theories, in which a simple global symmetry group acts irreducibly 
on \emph{all}
the vector fields $A_{\mu}^{\ti}$.

In general, the global symmetries of a 5D, $\mathcal{N}=2$
MESGT can be divided into two categories:
\begin{itemize}
\item Any $\mathcal{N}=2$ MESGT is always
invariant under the global R-symmetry
group $SU(2)_{R}$. As mentioned earlier, $SU(2)_{R}$  acts nontrivially only
on the fermions $\Psi_{\mu}^{i}$ and
$\lambda^{i\ta}$. In particular, it does not act on the vector fields
$A_{\mu}^{\ti}$.
\item Any group, $G$, of linear transformations
\begin{equation} \label{hAtrafo}
h^{\ti}\rightarrow {B^{\ti}}_{\tj}h^{\tj}, \quad A_{\mu}^{\ti}\rightarrow
{B^{\ti}}_{\tj}A_{\mu}^{\tj}
\end{equation} that leave the
tensor $C_{\ti\tj\tk}$ invariant
\begin{displaymath}
{B^{\ti'}}_{\ti}{B^{\tj'}}_{\tj}{B^{\tk'}}_{\tk}C_{\ti'\tj'\tk'}
=C_{\ti\tj\tk} \ ,
\end{displaymath}
is automatically a symmetry of the entire Lagrangian (\ref{Lagrange}),
since the latter
is uniquely  determined by the $C_{\ti\tj\tk}$.
These symmetries act as
isometries of the scalar manifold $\mathcal{M}$,
which becomes evident if one rewrites
the kinetic energy term for the scalar fields as \cite{GST1,dWvP1}
\begin{displaymath}
-\frac{1}{2}g_{\tx\ty}(\partial_{\mu}\varphi^{\tx})(\partial^{\mu}
\varphi^{\ty})= \frac{3}{2}C_{\ti\tj\tk}h^{\ti}\partial_{\mu}h^{\tj}
\partial^{\mu}h^{\tk} \ ,
\end{displaymath}
with the $h^{\ti}$ being constrained according to (\ref{hyper}).
\end{itemize}

As there is no interference between these two types of symmetries
(a consequence of the vector and scalar fields being $SU(2)_{R}$-inert),
 the full global symmetry group
 of (\ref{Lagrange})  factorizes into $SU(2)_{R}\times
G$. Obviously, then, any `unifying' symmetry group has to be a subgroup
of $G$, because only this group acts non-trivially on the vector fields.

For a generic MESGT, the invariance group $G$ of the underlying cubic
polynomial $\mathcal{V}(h)$ can be rather small or even trivial.
A well-studied class of theories with rather large symmetry groups $G$ are
the ones whose target spaces $\M$ are symmetric spaces. This class of
MESGTs can be divided into three families \cite{GST1,GST3}:

\begin{enumerate}
\item The ``generic'' or ``reducible'' Jordan family:\\
\eq \mathcal{M} = \
\frac{SO(\tn-1,1)}{SO(\tn-1)}\times SO(1,1), \qquad \tn\geq
1.\nonumber \en

\item The ``irreducible'' or ``magical'' Jordan family
\footnote{The name ``magical'' derives from the deep
connection with the ``magic square'' of Freudenthal,
Rosenfeld and Tits \cite{FRT}.} .
\begin{eqnarray}
\mathcal{M}&=& SL(3,\mathbb{R})/
SO(3)\qquad
(\tn=5)\nonumber\cr
\mathcal{M}&=& SL(3,\mathbb{C})/
SU(3)\qquad
(\tn=8)\nonumber\cr
\mathcal{M}&=& SU^{*}(6)/
Usp(6)\qquad
(\tn=14)\nonumber\cr
\mathcal{M}    &=& E_{6(-26)}/
F_{4}\qquad \qquad
(\tn=26)\nonumber
\end{eqnarray}

\item  The symmetric non-Jordan family:
\begin{equation}\mathcal{M}=\frac{SO(1,\tn)}{SO(\tn)},\quad \tn>1.
\end{equation}.
\end{enumerate}

The reason for the names of these three families will become clear
in the next section. Here, we   focus on the question which of the
above theories are unified MESGTs.
Let us first consider the generic Jordan family (i).
For this family, the isometry group of the scalar
manifold $\M$ is given by $SO(\tn -1,1)\times SO(1,1)$.
This is also the the symmetry group, $G$,  of the underlying cubic
polynomial
\begin{equation}\label{genJorpol}
\mathcal{V}(h)= \frac{3\sqrt{3}}{2}h^0 [(h^1)^2-(h^{2})^2-\ldots
-(h^{\tn})^2 ] \ ,
\end{equation}
where $SO(1,1)$ acts by rescalings $(h^0,h^1,\ldots,h^{\tn})  \rightarrow
 (\lambda^2 h^0, \lambda^{-1}h^1,\ldots,\lambda^{-1}h^{\tn})$,
and $(h^1,\ldots,h^{\tn})$ transform in the fundamental
representation of $SO(\tn-1,1)$. As $h^{0}$ is inert under
$SO(\tn-1,1)$, there can be no simple subgroup of $G$ under which
all $h^{\ti}$ (and thus all vector fields $A_{\mu}^{\ti}$)
transform irreducibly. Hence, the generic Jordan family does not
contain any unified MESGTs.

Let us now turn to the ``magical'' Jordan family (ii). For these
theories, the isometry groups of the scalar manifolds $\M$ are
$SL(3,\mathbb{R})$, $SL(3,\mathbb{C})$, $SU^{\ast}(6)$ and
$E_{6(-26)}$, respectively. Just as in the generic Jordan family (i),
these are also the symmetry groups $G$ of the underlying cubic
polynomials $\mathcal{V}(h)$. Under these simple symmetry groups
$G$, the, respectively, $6$, $9$, $15$ and $27$ vector fields
$A_{\mu}^{\ti}$ transform irreducibly \cite{GST1}. Thus, according to our
definition, all four theories of the magical Jordan family are
unified MESGTs.

For the two families we have discussed so far, the symmetry group
$G$ of the $C_{\ti\tj\tk}$ always coincided with
the full isometry group of the scalar manifold $\mathcal{M}$.
For the third family (iii) of symmetric spaces, i.e. for the symmetric
non-Jordan family, this is no longer true \cite{dWvP2}:
Whereas the isometry group of $\M$ is
$SO(1,\tn)$,
the symmetry group $G$ of the $C_{\ti\tj\tk}$ (i.e., the symmetry
group of the whole Lagrangian) is only
the subgroup $E_{(\tn -1)} \times SO(1,1)$, where  $E_{(\tn -1)}$
denotes the Euclidean group in $(\tn -1)$ dimensions,
\[
E_{(\tn -1)} =SO(\tn-1) \ltimes T_{(\tn-1)} \ ,
\]
where $T_{(\tn -1)}$
is the group of translations in an $(\tn -1)$ dimensional
Euclidean space \cite{dWvP2}, and  $\ltimes$ denotes the semi-direct product.
A simple subgroup of this group has to
be a subgroup of $SO(\tn-1)$, under which only $(\tn-1)$ of the $(\tn+1)$
vector fields transform nontrivially \cite{dWvP2}. Hence, the symmetric
non-Jordan family does not provide us with any new unified MESGTs.

To sum up, of all  the 5D, $\mathcal{N}=2$ MESGTs whose
scalar manifolds are symmetric spaces only those of
the magical Jordan family (ii) are {\it unified} MESGTs.
This statement can even be extended to the larger class of theories in
which $\M$ is  \emph{homogeneous}, but not necessarily symmetric.
The possible homogeneous scalar manifolds were classified in
\cite{dWvP1}, and it is easy to see from the list given in
\cite{dWvP1}  that also in that class the only possible unified MESGTs are
provided by the four magical theories described above.

The goal of the first part of this paper is to find more examples
of unified MESGTs by abandoning the restriction that $\M$ be a
symmetric or a homogeneous space\footnote{In fact, there is no good
physical motivation for such a restriction.
For one thing, the quantum corrected low energy effective actions of
$\mathcal{N}=2 $ compactifications of string  or M-theory are in
general not based on homogeneous target spaces (see e.g. \cite{AFT,MZ} 
for
some explicit 5D examples).
Moreover, even in an intrinsically 5D  framework, the
deviation from the class of homogeneous target  spaces is often
a crucial step towards obtaining models with interesting physical
properties   (see e.g.    \cite{BehrndtDallAgata, EGZ, DGKL}).}. The
mathematical problem one has to solve in order to find such
theories is easily stated: Find irreducible representations of
simple groups $G$ with an invariant third rank symmetric tensor
$C_{\ti\tj\tk}$ such that the resulting metrics $g_{\tx\ty}$ and
${\stackrel{\circ}{a}}_{\ti\tj}$ are positive definite, at least
in
 the vicinity
of some point, $c$, on the hypersurface
$C_{\ti\tj\tk}h^{\ti}h^{\tj}h^{\tk}=1$. An equivalent way of stating
the positivity properties of $g_{\tx\ty}$ and
${\stackrel{\circ}{a}}_{\ti\tj}$ is to require that the $C_{\ti\tj\tk}$
can be brought to the canonical form (\ref{canbasis}) by means of a linear
redefinition of the $h^{\tj}$.

In the following section, we show that an infinite number of solutions to this
mathematical problem can be easily constructed using
the language of Jordan algebras.



\section{Jordan algebras and unified MESGTs}
\setcounter{equation}{0}
As we have seen in the previous section, within the class of
symmetric or homogeneous scalar manifolds only four give rise to a
unified MESGT. These four theories are all members of what we
called the ``magical Jordan family'' (family (ii)). The magical
Jordan family and the generic Jordan family (family (i)) owe their
names to the fact that they are associated with Jordan algebras.
The ``symmetric non-Jordan family'' (family (iii)), by contrast,
is, as the name suggests,  not connected to Jordan algebras. In
order to become more explicit, let us first recall the definition
of a Jordan algebra.\\
\vspace{-1mm}\\
\textbf{Definition 1:}
 A Jordan algebra over a field $\mathbb{F}$ (which we take to be 
$\mathbb{R}$ or
$\mathbb{C}$) is an  algebra, $J$, over $\mathbb{F}$ 
with a symmetric product $\circ$,
\begin{equation}\label{commute} X\circ Y = Y
\circ X \in J, \quad \forall\,\, X,Y \in J \ ,
\end{equation}
that satisfies
the Jordan identity
\begin{equation}\label{Jidentity}
X\circ (Y \circ X^2)= (X\circ Y) \circ X^2 \ ,
\end{equation}
where $X^2\equiv (X\circ X)$.
\vspace{-1mm}\\

The Jordan identity (\ref{Jidentity}) is automatically satisfied when
the product $\circ$ is associative, but
(\ref{Jidentity}) does not imply associativity. In other words, a
Jordan algebra is commutative, but in general not associative.
Historically, Jordan algebras were introduced in an attempt to
generalize the formalism of quantum mechanics by
capturing the algebraic essence of Hermitian operators corresponding
to observables without reference to the
underlying Hilbert space on which they act \cite{JvNW}. While the
Hermitian    operators acting on a Hilbert
space do not close under the ordinary (associative) operator product,
they do close and form a Jordan algebra
under the Jordan product $\circ$ defined as one half the
anticommutator \footnote{The axioms (\ref{commute}) and
(\ref{Jidentity}) are of course also fulfilled for other multiples of the
anticommutator. The prefactor one half
is singled out as the only prefactor for which eq. (\ref{Norm})
becomes independent of the degree $p$ of the
Jordan algebra.}.

For every Jordan algebra $J$, one can define a norm
form, $N:J\rightarrow \mathbb{R}$, that satisfies the
composition property \cite{jacobson}
\begin{equation}\label{Norm}
N[2X\circ(Y\circ X)-(X\circ X)\circ Y]=N^{2}(X)N(Y).
\end{equation}
The degree, $p$, of the norm form is defined by $N(\lambda X)=\lambda^p N(X)$,
where $\lambda\in \mathbb{R}$. $p$ is also called
 the degree of the Jordan algebra.\\
\vspace{-1mm}\\
\textbf{Definition 2:}
A \emph{Euclidean} Jordan algebra is a Jordan algebra for which
the condition $X\circ X + Y\circ Y=0$ implies that $X=Y=0$ for all
$X,Y\in J$. The automorphism groups of Euclidean Jordan algebras
are always compact.\\
\vspace{-1mm}\\
In \cite{GST1}, it was shown, that whenever a Jordan algebra is
Euclidean, and its norm form $N$ is cubic ($p=3$), one can identify the
norm form $N$ with the cubic polynomial $\mathcal{V}$ of a MESGT
so that
 $g_{\tx\ty}$ and ${\stackrel{\circ}{a}}_{\ti\tj}$ are positive definite.
The MESGTs whose cubic polynomial arise in this way, are precisely the
first two families (i) and (ii). The relevant Jordan algebras
are
\begin{enumerate}
\item $J=\mathbb{R}\oplus \Sigma_{\tn}$ for the generic
Jordan family with the scalar manifolds
\[ \mathcal{M} =
\frac{SO(\tilde{n}-1,1)}{SO(\tilde{n}-1)}\times
SO(1,1) \ .
\] Here, $\Sigma_{\tn}$ is a Jordan algebra of degree $p=2$
associated with a quadratic norm form in $\tn$ dimensions that
has a ``Minkowskian signature'' $(+,-,\ldots,-)$.
A simple realization of $\Sigma_{\tn}$ is provided by $(\tn-1)$ Dirac
gamma matrices $\gamma^i$
$(i,j,\ldots=1,\ldots,(\tn-1))$ of an $(\tn-1)$ dimensional Euclidean
space together with the identity matrix $
\gamma^0 = \mathbf{1}$  and the Jordan product $\circ$ being one half the
anticommutator:

\begin{eqnarray}
\gamma^i \circ \gamma^j &=& \frac{1}{2}
\{\gamma^i,\gamma^j\}= \delta^{ij} \mathbf{1}
\nonumber \\
\gamma^0 \circ \gamma^0 &=& \frac{1}{2}
\{\gamma^0,\gamma^0\}= \mathbf{1}\nonumber\\
\gamma^i \circ \gamma^0 &=& \frac{1}{2}
\{\gamma^i,\gamma^0\}= \gamma^i \ .
\end{eqnarray}
  The norm of a general element $X = X_0 \gamma^0 + X_i \gamma^i $ of
$\Sigma_{\tn}$
is defined as
\[  N(X) = \frac{1}{2^{[\tn/2]}} \textrm{Tr}
X \bar{X} = X_0X_0 - X_iX_i \ ,  \] where
\[ \bar{X} \equiv  X_0 \gamma^0 - X_i \gamma^i  \ . \] 
 The norm of a general element $y \oplus X $ of the non-simple
Jordan algebra $J=\mathbb{R}\oplus \Sigma_{\tn}$ is
simply given by $y N(X)$ (cf eq. (\ref{genJorpol})).
 \item The magical
Jordan family corresponds to the four \emph{simple} Euclidean Jordan algebras
of degree 3. These simple Jordan algebras are denoted by $J_3^{\mathbb{R}}$,
$J_3^{\mathbb{C}}$, $J_3^{\mathbb{H}}$, $J_3^{\mathbb{O}}$ and 
are isomorphic to the Hermitian $(3\times 3)$-matrices over the four
division algebras $\mathbb{R}, \mathbb{C},
\mathbb{H}, \mathbb{O}$ with the product being one half the anticommutator:
\begin{eqnarray}\label{magicals}
J_{3}^{\mathbb{R}}:\quad \mathcal{M}&=& SL(3,\mathbb{R})/
SO(3)\qquad
(\tn=5)\nonumber\\
J_{3}^{\mathbb{C}}:\quad \mathcal{M}&=& SL(3,\mathbb{C})/
SU(3)\qquad
(\tn=8)\nonumber\\
J_{3}^{\mathbb{H}}:\quad \mathcal{M}&=& SU^{*}(6)/
Usp(6)\qquad
(\tn=14) \nonumber \\
J_{3}^{\mathbb{O}}:\quad \mathcal{M}&=& E_{6(-26)}/
F_{4}\qquad \qquad
(\tn=26) \ .
\end{eqnarray}
The cubic norm form, $N$, of these Jordan algebras is given by the 
determinant  of the corresponding Hermitian $(3\times 3)$-matrices.
\end{enumerate}

In all the above examples, the scalar manifold is given by
 $\mathcal{M}= \frac{\textrm{Str}_{0}(J)}{\textrm{Aut}
(J)}$, where
$\textrm{Str}_{0}(J)$ and $\textrm{Aut}(J)$  are, respectively,
 the reduced structure group\footnote{The reduced structure group
$\textrm{Str}_{0}(J)$ is simply
the invariance group of the norm form, $N$, of the corresponding
Jordan algebra $J$. As such, it is, for the above Jordan algebras,
 isomorphic to the symmetry group   $G$  of the
corresponding MESGTs.}
and the automorphism
group of the corresponding Jordan algebra $J$
\cite{GST1, KMC}.


\subsection{The novel families of unified MESGTs
defined by simple non-compact Jordan algebras}

In this section we shall investigate the question whether there
exist unified $\mathcal{N}=2$ MESGTs beyond the four magical ones. We
find that there do indeed exist three novel infinite families of
unified $\mathcal{N}=2$ MESGTs, as well as another exceptional unified
 MESGT beyond the
magical theories. The three infinite families are associated
with Jordan algebras of arbitrary high degree $p\geq 3$ that  are no
longer Euclidean.

Let us first try to understand heuristically why Jordan algebras
also play a natural r\^{o}le
in these novel families of unified
MESGTs. In the two Jordan families (i) and (ii) discussed in
 the previous section,
 the cubic polynomial $\mathcal{V}$ defined by the
symmetric tensor $C_{\ti\tj\tk}$ of the supergravity theory
is identified with the norm form, $N$,
of a Euclidean Jordan algebra of degree three. As a consequence, the
invariance group of the norm form  becomes a symmetry group, $G$,  of the
supergravity Lagrangian.
Clearly,
the restriction to Jordan algebras of degree  three
is crucial for this identification, because  for Jordan algebras of degree
$p>3$ the norm forms are, by definition,
 no longer cubic. So, if we are   to find new unified MESGTs,
we will certainly not find them  by identifying the cubic polynomial
$\mathcal{V}$ with norm forms of other Jordan algebras.
All the cases where a norm form of a Jordan algebra can be identified
with an admissible supergravity polynomial $\mathcal{V}$ are already
exhausted by the two families we discussed in the previous section.

In order to find new unified MESGTs, one would therefore have to
identify the $C_{\ti\tj\tk}$ of such a  MESGT
with another mathematical object that admits the  action of a
non-trivial invariance group $G$. As a rather
natural object of this sort, one could try the structure constants of
an algebra. The fact that the
$C_{\ti\tj\tk}$ are completely symmetric in three indices
 implies that if they  are
to be identified with the structure constants of some algebra,
that algebra must have a symmetric product. Jordan
algebras are, of course, some of the best known and studied algebras
with a symmetric product. Furthermore,
Jordan algebras are the natural mathematical structures that arise in
the study of domains of positivity
\cite{positivitydomain},
 which in our case are related
to the positivity of the kinetic energy terms of the scalar and
vector fields.

We are thus led to investigate the possibility of identifying the
\emph{structure constants} of Jordan algebras
with the constants $C_{\ti\tj\tk}$ of the supergravity theory.
However,  if we are   to identify the
$C_{\ti\tj\tk}$ with the full set of structure constants of a
simple Jordan algebra
 then the corresponding $\CN=2$ MESGT
can not be a \emph{unified} theory since the invariance group of the 
structure constants is the automorphism
group, under which the identity element of a simple Jordan algebra is 
a singlet. Because of this singlet, the
automorphism group would then not act irreducibly on all the vector 
fields of the theory. Therefore, to be able
to obtain a \emph{unified} MESGT, we can at best try to identify 
$C_{\ti\tj\tk}$ with a subset of the structure
constants that does not involve the identity element.

Furthermore, from the general form of the constants $C_{\ti\tj\tk}$
in the canonical basis  we expect any symmetry group under which all
the vectors transform in a single irreducible representation
 to be non-compact \cite{GST1,dWvP2,EGZ}.
Thus we
are led to investigate Jordan algebras with non-compact automorphism
groups.

The Jordan algebras of $(n\times n)$ Hermitian matrices over
various division algebras are  Euclidean (compact) Jordan
algebras whose automorphism groups  are compact. In particular, the Jordan
algebras of degree 3 discussed in the previous section are all Euclidean.
Non-compact analogs
$J_{(q,n-q)}^{\mathbb{A}} $ of Euclidean Jordan algebras
$J_n^{\mathbb{A}}$ of $(n\times n)$ Hermitian matrices over
the associative division
algebras $\mathbb{A}= \mathbb{R},\mathbb{C}, \mathbb{H} $
for $n \geq 3$ and of  $J_3^{ \mathbb{O}}$  \footnote{The Hermitian
$(n\times n)$ matrices
over the octonions do not form Jordan algebras for $n\neq 3$.}
 are realized by
matrices that are Hermitian with respect to a non-Euclidean
``metric'' $\eta $ with signature $(q,n-q)$:
\begin{equation}\label{eta}
(\eta X)^\dag = \eta X  \hspace{1cm} \forall  X\in J_{(q,n-q)}^{\mathbb{A}}
\ .
\end{equation}
Obviously, if we choose $\eta$ to have a Euclidean signature,
we obtain back Euclidean (compact)
Jordan algebras. Consider now Minkowskian Jordan
algebras $J_{(1,N)}^{\mathbb{A}} $
of degree $n=N+1$ defined by choosing $\eta$ to be the
Minkowski metric $\eta = (-,+,+,...,+)$. A general element, $U$, of
$J_{(1,N)}^{\mathbb{A}} $ can be written in the form

\begin{equation}
U=\left(\begin{array}{cc}
x & -Y^{\dagger}\\
Y & Z
\end{array}\right),
\end{equation}
where $Z$ is an element of the Euclidean subalgebra
$J_{N}^{\mathbb{A}} $ (i.e., it is a Hermitian $(N\times N)$-matrix over 
$\mathbb{A}$), $x\in\mathbb{R}$, and $Y$ denotes an
$N$-dimensional  column vector over $\mathbb{A}$.
Under the automorphism group, $\textrm{Aut}(J_{(1,N)}^{\mathbb{A}})$,
the simple Jordan algebras $J_{(1,N)}^{\mathbb{A}}$ decompose into an
irreducible representation formed by the traceless elements
plus a singlet, which is given by the identity element of
$J_{(1,N)}^{\mathbb{A}}$
(i.e., by the unit matrix $U=\mathbf{1}$):
\begin{equation}
J_{(1,N)}^{\mathbb{A}}= \mathbf{1}\oplus \{ \textrm{traceless elements}\} \ .
\end{equation}
The traceless elements do not close under the Jordan product, $\circ$,
but one can define a symmetric product, $\star$, under
 which
the traceless elements close as follows: \footnote { Such a product was
introduced among the Hermitian generators
of $SU(N)$ by Michel and Radicati sometime ago \cite{mira}.
 Note that Hermitian generators of $SU(N)$ are in
one-to-one correspondence with the traceless elements of
 $J_N^{\mathbb{C}}$.  The ``symmetric'' algebras with the star product $\star$
do not have an identity element.}

\[ A \star B := A \circ B - \frac{1}{(N+1)} \textrm{tr}(A \circ B)
\mathbf{1} \ ,
\]
where $\circ$ is the Jordan product \[ A \circ B = \frac{1}{2} (
AB+BA)\ .  \]
 Thus, the structure constants ($d$-symbols)  of the traceless elements
under the symmetric $ \star $ product will be invariant tensors of
the automorphism groups $\textrm{Aut}(J_{(1,N)}^{\mathbb{A}})$
of the Jordan algebras. Denoting
the traceless elements as
$T_{\ti}$ $(\ti=0,\ldots, (D-2))$ with $D$ being the dimension of
$J_{(1,N)}^{\mathbb{A}} $, we have
\[
T_{\ti} \star T_{\tj} =d_{\ti\tj}^{~~~\tk} T_{\tk} \ .
\]
The $d$-symbols are then given by

\begin{equation}\label{dsymbols}
d_{\ti\tj\tk}\equiv d_{\tj\tk}^{~~~\tl} \tau_{\tl\ti} = \frac{1}{2}
\textrm{tr}(T_{\ti}\{ T_{\tj}, T_{\tk}\})=  
\textrm{tr} (T_{\ti}\circ (T_{\tj}\circ
T_{\tk})) 
\end{equation}
where \footnote{One can choose the elements $T_{\ti}$ such that 
$\tau_{\ti\tj} = \delta_{\ti\tj} $ 
($\tau_{\ti\tj} = - \delta_{\ti\tj} $)  for two compact (noncompact) 
elements $ T_{\ti}$ and $T_{\tj}$ and zero
otherwise.}
\[\tau_{\tl\ti} = \textrm{tr} (T_{\tl}\circ T_{\ti}) \ . \]  
The $d_{\ti\tj\tk}$ are completely symmetric in their indices, and as
$\textrm{Aut}(J_{(1,N)}^{\mathbb{A}})$ acts irreducibly on the traceless
elements $T_{\ti}$, the $d_{\ti\tj\tk}$ are a promising candidate
for the $C_{\ti\tj\tk}$ of a unified MESGT. What remains to check, however,
 is
whether the metrics $g_{\tx\ty}$ and ${\stackrel{\circ}{a}}_{\ti\tj}$ on the
resulting scalar manifold $\M$ are really positive definite. As we will see,
this is true if and only if the signature of $\eta$ is really $(+,-,\ldots,-)$.

Let us therefore now assume that $C_{\ti\tj\tk}= d_{\ti\tj\tk}$ defines a
cubic polynomial $\mathcal{V}(h)=d_{\ti\tj\tk} h^{\ti} h^{\tj} h^{\tk}$
of an $\mathcal{N}=2$ MESGT. As explained in Section 2, the
$\tn$-dimensional scalar manifold $\M$ is given by the hypersurface
$\mathcal{V}(h)=1$ in the auxiliary space $\mathbb{R}^{(\tn+1)}$
spanned by the $h^{\ti}$. This auxiliary space $\mathbb{R}^{(\tn+1)}$
can be identified with the traceless subspace, $J_{0(1,N)}^{\mathbb{A}}$,
of the Jordan algebra $J_{(1,N)}^{\mathbb{A}}$ (i.e., the dimension, $D$,
of $J_{(1,N)}^{\mathbb{A}}$ and $\tn$ are related by $D-1=\tn+1$).
We will now show that a judicious choice of the traceless generators
$T_{\ti}$ will bring the $d$-symbols $d_{\ti\tj\tk}$ (eq. (\ref{dsymbols}))
into the canonical form (\ref{canbasis}). This demonstrates the
positivity of the resulting metrics $g_{\tx\ty}$ and
${\stackrel{\circ}{a}}_{\ti\tj}$.

We start by noting that any point, c, on the scalar manifold
$\ensuremath{\mathcal{M}}$ defines a non-zero
 element $ h^{\ensuremath{\tilde{I}}}(c)$ in the
embedding space $\mathbb{R}^{(\tn+1)}$
(it has to be non-zero, because $h^{\ensuremath{\tilde{I}}}(c)=0$ would
be inconsistent with $\mathcal{V}|_{\ensuremath{\mathcal{M}}}=1$), and thus a
non-trivial direction in the  traceless subspace $J_{0(1,N)}^{\mathbb{A}}$
of the Jordan
algebra $J_{(1,N)}^{\mathbb{A}}$. We choose our coordinates
$h^{\ensuremath{\tilde{I}}
}$ such that $h^{\ensuremath{\tilde{I}}}(c)=(1,0,\ldots ,0)$, and,
correspondingly, the generators $T_{\ensuremath{\tilde{I}}}$ such
that $T_{0}$ is aligned with the non-trivial direction defined
by $h^{\ensuremath{\tilde{I}}}(c)$. Note that
\begin{equation}
1=\mathcal{V}(h(c))=d_{\ti\tj\tk}h^{\ti}(c)h^{\tj}(c)h^{\tk}(c)=d_{000} \ .
\end{equation}
A general traceless element $T_{\ti}$ in  $J_{(1,N)}^{\mathbb{A}}$ can be
represented as
\begin{equation}
T_{\ti}=\left(
\begin{array}{cc}
x & -Y^{\dagger } \\
Y & Z
\end{array}
\right) ,
\end{equation}
where $x\in \mathbb{R}$, $Y$ denotes an $N$-dimensional column vector over
$\mathbb{A}$, $Y^{\dagger }$ its Hermitian conjugate  and $Z$ is a Hermitian $
(N\times N)$ matrix over $\mathbb{A}$  with $\textrm{tr}(Z)=-x$.
We choose
\begin{equation}
T_{0}=\left(
\begin{array}{cc}
a & 0 \\
0 & -\frac{a}{N}\mathbf{1}_{(N)}%
\end{array}
\right) ,
\end{equation}
where $a$ is some real number fixed to be
\begin{equation}\label{a>0}
a=\left( \frac{N^{2}}{(N^2-1)}\right) ^{\frac{1}{3}}>0
\end{equation}
by the condition  $d_{000}=1$.
We use $T_{M}, T_{N}\ldots$ to denote generic elements of the form
\begin{equation}
T_{M}=\left(
\begin{array}{cc}
0 & -Y^{\dagger} \\
Y & 0
\end{array}
\right) ,
\end{equation}
and $T_A,T_B,\ldots$ for the generators of the type
\begin{equation}
T_{A}=\left(
\begin{array}{cc}
0 & 0 \\
0 & Z
\end{array}
\right) ,
\end{equation}
with $\textrm{tr}(Z)=0$.
It is easy to see that the $d$-symbols of the type $d_{00M}$ and $d_{00A}$
vanish:
\begin{equation}
d_{00M}=d_{00A}=0 \ .
\end{equation}
Obviously, the $d$-symbols of the type $d_{0MN}$, $d_{0MA}$ and $d_{0AB}$
define a metric on the subspace spanned by the $h^{M}$ and $h^{A}$.
 This metric is negative definite, as one easily confirms by calculating 
the diagonal elements (no sum):
\begin{eqnarray}
d_{0MM} &=&- a(Y^{\dagger }Y)\left( 1-\frac{1}{N}\right) <0 \\
d_{0AA} &=&-\frac{a}{N}\textrm{tr}(Z^2) <0 \ ,
\end{eqnarray}
where $a>0$ (eq. (\ref{a>0})) has been used.
As $d_{0MA} =0$, one can always go to a basis such that
\begin{eqnarray}
d_{0MN} &=&-\frac{1}{2}\delta_{MN}\\
d_{0AB} &=&-\frac{1}{2}\delta_{AB}\\
d_{0MA} &=&0 \ .
\end{eqnarray}
Together with $d_{000}=1$, $d_{00M}=d_{00A}=0$, this implies that the
polynomial $\mathcal{V}(h)=d_{\ti\tj\tk}h^{\ti} h^{\tj} h^{\tk}$ can
always be brought to the canonical form (\ref{canbasis}) if one identifies
$\ti=(0,i)=(0,M,A)$. The metrics $g_{\tx\ty}$ and
${\stackrel{\circ}{a}}_{\ti\tj}$ are thus positive definite, at least in the
vicinity of the base point $c$. Note that this positivity requirement
is precisely the point where
the Minkowski signature $(1,N)$ of the metric $\eta$ (cf. eq. (\ref{eta}))
becomes important. For metrics $\eta$ with non-Minkowskian signature,
the diagonal elements
 $d_{0MM}$ and $d_{0AA}$ would not all be negative.
Hence, only the Minkowskian signatures can lead to physically
acceptable unified MESGTs.

Putting everything together, we have thus shown the following:
If one identifies the $d$-symbols (\ref{dsymbols}) of the traceless elements
of a Minkowskian Jordan algebra $J_{(1,N)}^{\mathbb{A}}$ with the
$C_{\ti\tj\tk}$ of a MESGT: $C_{\ti\tj\tk}=d_{\ti\tj\tk}$,
one obtains a unified MESGT, in which all the vector fields transform
irreducibly under the simple automorphism group
$\textrm{Aut}(J_{(1,N)}^{\mathbb{A}})$ of that Jordan algebra.

For $\mathbb{A}=\mathbb{R},\mathbb{C},\mathbb{H}$ one obtains in
this way three infinite families of physically
acceptable unified MESGTs (one for each $N\geq2$). For the
octonionic case, the situation is a bit different. The
$d$-symbols of the octonionic Minkowskian Hermitian $(N+1)\times
(N+1)$-matrices with the anticommutator product
all lead to positive definite metrics $g_{\tx\ty}$ and
${\stackrel{\circ}{a}}_{\ti\tj}$, i.e., to physically
acceptable MESGTs. For $N\neq  2$, however, these octonionic
Hermitian matrix algebras are no longer Jordan
algebras.
 Surprisingly, the automorphism groups of these octonionic  algebras
  for $N\geq 3$ do not have the automorphism group
 $F_{4(-20)}$ of $J_{(1,2)}^{\mathbb O}$ as a subgroup \cite{goodaire}.
Instead, the  automorphism groups for $N \geq 3$
 are direct product groups of the form  $SO(N,1) \times G_2$.  None
of these two factors acts irreducibly on all the traceless elements, and hence
  the corresponding $\mathcal{N}=2$ MESGTs are  not unified
 theories. Thus, the $\CN=2$ MESGT defined by the exceptional Minkowskian
Jordan algebra $J_{(1,2)}^{\mathbb{O}}$ is the only \emph{unified}
MESGT of this infinite tower of otherwise acceptable octonionic theories.

All these results are summarized in Table 1,
which  lists all the simple Minkowskian Jordan algebras of
type $J_{(1,N)}^{\mathbb{A}} $,
their  automorphism  groups and the numbers of vector and scalar
fields in the unified
MESGT's defined by them.

 \begin{table}[ht]

\begin{center}
\begin{displaymath}
\begin{array}{|c|c|c|c|c|}
\hline
~&~&~&~&~\\
J & D&  \textrm{Aut}(J)
& \textrm{No. of vector fields} & \textrm{No.  of scalars}  \\
\hline
~&~&~&~&~\\
J_{(1,N)}^{\mathbb R}& \frac{1}{2} (N+1)(N+2) & SO(N,1) &
\frac{1}{2}N(N+3) &\frac{1}{2}N(N+3)-1 \\
~&~&~&~&~\\
J_{(1,N)}^{\mathbb C} & (N+1)^2 & SU(N,1) & N(N+2) & N(N+2)-1 \\
~&~&~&~&~\\
J_{(1,N)}^{\mathbb H} &(N+1)(2N+1) & USp(2N,2) & N(2N+3) & N(2N+3)-1 \\
~&~&~&~&~\\
J_{(1,2)}^{\mathbb O} & 27& F_{4(-20)} & 26 & 25 \\
~&~&~&~&~ \\
\hline
\end{array}
\end{displaymath}
\end{center}

\caption{List of the simple Minkowskian Jordan algebras
of type $J_{(1,N)}^{\mathbb{A}}$.
 The columns show, respectively, their  dimensions $D$,
their automorphism groups $\textrm{Aut}
(J_{(1,N)}^{\mathbb{A}})$,
the number of vector fields $(\tn+1)=(D-1)$
and the number of scalars $\tn=(D-2)$ in the corresponding MESGTs}
\end{table}

As an interesting observation, one notes that the number of vector fields
for the theories defined by
$J_{(1,3)}^{\mathbb R}$, $J_{(1,3)}^{\mathbb C}$ and $J_{(1,3)}^{\mathbb H}$
are given by 9, 15 and 27, respectively. These are exactly the same
numbers of vector fields one finds in the magical theories based on the
norm forms of the Euclidean Jordan algebras
$J_{3}^{\mathbb C}$, $J_{3}^{\mathbb H}$ and $J_{3}^{\mathbb O}$, respectively.
As we will show in Section 6, this is not an accident: The magical
MESGTs based
on $J_{3}^{\mathbb C}$, $J_{3}^{\mathbb H}$ and $J_{3}^{\mathbb O}$ found in
\cite{GST1}
are \emph{equivalent} (i.e. the cubic polynomials $\mathcal{V}(h)$ agree)
to the ones we constructed in this paper using
the Minkowskian algebras
$J_{(1,3)}^{\mathbb R}$, $J_{(1,3)}^{\mathbb C}$ and $J_{(1,3)}^{\mathbb H}$,
respectively. This is related to
 a construction of the degree 3 simple Jordan algebras 
$J_{3}^{\mathbb C}$, $J_{3}^{\mathbb H}$ and $J_{3}^{\mathbb O}$
 in terms of the traceless elements of degree 
four simple Jordan algebras over $\mathbb{R}, \mathbb{C}$
and $\mathbb{H}$ by Allison and Faulkner \cite{alfa}.  
This implies, that the only known unified MESGT that is
not covered by the above table, is the magical theory of 
\cite{GST1} based on the Jordan algebra
$J_{3}^{\mathbb{R}}$ with $(\tn+1)=6$ vector fields and 
the target space $\M=SL(3,\mathbb{R})/SO(3)$ (see Section
3).


\section{Unified $\CN=2$  Yang-Mills-Einstein    supergravity theories}
\setcounter{equation}{0}


In the previous section, we have constructed infinitely many
novel unified  $\CN=2$ MESGTs in five dimensions by establishing a
relation to a
certain class of non-compact Jordan algebras.
As we did not give a completeness proof, it   is not clear
whether these novel theories
(together with the remaining magical one based on the compact Jordan
algebra $J^{\mathbb{R}}_{3}$ found in \cite{GST1})
exhaust all possible unified MESGTs in five dimensions.
In order to answer that question, one would have to show that there are
no further irreducible representations of simple groups with an invariant
symmetric tensor of rank three that gives rise to positive metrics
$g_{\tx\ty}$ and
${\stackrel{\circ}{a}}_{\ti\tj}$. We leave this as an open problem.

Instead, we will now, in this section,
try to construct novel  unified \emph{Yang-Mills}-Einstein
supergravity theories, i.e., theories in which
 all the vector fields, including the graviphoton, transform irreducibly
in the adjoint representation of a simple \emph{local} gauge group, $K$.
If one turns off the gauge coupling of such a unified YMESGT, the local
symmetry group $K$ becomes a global symmetry group under which the vector
fields still transform irreducibly. In other words, turning off the
gauge coupling of a unified YMESGT yields a unified MESGT.
Conversely, any unified YMESGT
can be obtained from a unified MESGT, by gauging a suitable subgroup
$K\subset G $ of the global symmetry group $G$ of the MESGT.

Let us briefly review some of the technical aspects of such
a gauging \cite{GST2,GST4}. In the unified MESGT one starts with,
the $(\tn+1)$ vector
fields $A_{\mu}^{\ti}$ form an irreducible $(\tn+1)$-dimensional
representation
of the simple global symmetry group $G$. If one wants to construct
a unified YMESGT out of this unified MESGT,
one has to gauge an $(\tn+1)$-dimensional
simple subgroup $K$ of $G$. For this to be
possible, the $(\tn+1)$-dimensional
  representation of $G$ has to reduce
to the adjoint representation of $K$ under the restriction $G\rightarrow K$:
\begin{equation}
\mathbf{(\tn+1)_{G} }\rightarrow   \textrm{adjoint}(K) \ .
\end{equation}
The only fields in the $\CN=2$ MESGT that transform nontrivially
 under $K$ are
the scalar fields $\varphi^{\tx}$, the spinor fields
$\lambda^{i\ta}$ and the vector fields $A_{\mu}^{\ti}$, ($\ti=1,\ldots,
\textrm{dim} K$). The $K$-covariantization is then achieved by
first replacing the corresponding derivatives/field strengths by
their $K$-gauge covariant counterparts:
\begin{eqnarray}
\partial_{\mu}\varphi^{\tx}&\longrightarrow & \mathcal{D}_{\mu}
\varphi^{\tx}\equiv
\partial_{\mu}\varphi^{\tx}+gA_{\mu}^{\ti}K_{\ti}^{\tx}\nonumber\\
\nabla_{\mu}\lambda^{i\ta}&\longrightarrow & \mathcal{D}_{\mu}
\lambda^{i\ta}\equiv \nabla_{\mu}\lambda^{i\ta} +gA_{\mu}^{\ti}
L_{\ti}^{\ta\tb}\lambda^{i\tb}
\nonumber\\
F_{\mu\nu}^{\ti}&\longrightarrow &\mathcal{F}_{\mu\nu}^{\ti}\equiv
F_{\mu\nu}^{\ti}+gf_{\tj\tk}^{\ti}A_{\mu}^{\tj}A_{\nu}^{\tk}.
\end{eqnarray}
Here,  $g$ denotes the coupling constant of $K$, $K_{\ti}^{\tx}$ are
the Killing vectors that generate the
subgroup $K \subset G$ of
isometries of the scalar manifold  $\mathcal{M}$ (cf. \cite{GST2}),
$L_{\ti}^{\ta\tb}$ are the (scalar field dependent) $K$-transformation
matrices of the fermions $\lambda^{i\ta}$ (cf. \cite{GST2,GST4}),
and $ f_{\ti\tj}^{\tk}$ are the structure constants of $K$.
The proper gauge-covariantization of the $F\wedge F\wedge A$-term
in (\ref{Lagrange}) leads to a Chern Simons term, i.e.,
\[\frac{e^{-1}}{6\sqrt{6}}
C_{\ti\tj\tk}\varepsilon^{\mu\nu\rho\sigma\lambda}
 F_{\mu\nu}^{\ti}F_{\rho\sigma}^{\tj}A_{\lambda}^{\tk}\] has to be
 replaced by
\begin{eqnarray}\label{CSForm}
\frac{e^{-1}}{6\sqrt{6}}C_{\ti\tj\tk}\varepsilon^{\mu\nu\rho\sigma\lambda}
\left\{ F_{\mu\nu}^{\ti}F_{\rho\sigma}^{\tj}A_{\lambda}^{\tk} +
\frac{3}{2}g
F_{\mu\nu}^{\ti}A_{\rho}^{\tj}(f_{\tilde{L}\tilde{M}}^{\tk}
A_{\sigma}^{\tilde{L}}A_{\lambda}^{\tilde{M}}) \right. +\left.
\frac{3}{5}g^{2}(f_{\tilde{N}\tilde{P}}^{ \tj}
A_{\nu}^{\tilde{N}}A_{\rho}^{\tilde{P}})
(f_{\tilde{L}\tilde{M}}^{\tk}A_{\sigma}^{\tilde{L}}A_{\lambda}^{\tilde{M}})
A_{\mu}^{\ti}\right\}.
\end{eqnarray}
Supersymmetry is broken by these replacements. In order to restore it,
one has to add a Yukawa-like term to the (covariantized) Lagrangian
\cite{GST2,GST4}:
\begin{equation}\label{Yukawaterm}
\mathcal{L}'=
-\frac{i}{2}g{\bar{\lambda}}^{i\ta}\lambda_{i}^{\tb}K_{\ti[\ta}
h^{\ti}_{\tb]} \ ,
\end{equation}
where $h^{\ti}_{\tb}$ is  essentially the derivative of $h^{\ti}$ with respect
to the scalar fields $\varphi^{\tx}$ (see \cite{GST1} for details).
The covariantized supersymmetry transformation laws remain unmodified.
Note that (in the absence of tensor multiplets)
such  an $\CN=2$ Yang-Mills-Einstein supergravity theory
has no scalar potential, i.e., the ground states are Minkowski space-times.


\subsection{The complete list
of 5D,   $\CN=2$ unified Yang-Mills-Einstein
  supergravity theories}
While we did not give a completeness proof of our list of unified MESGTs,
we are able to give a complete list of all possible
   unified  $\CN=2$ YMESGTs in
 five dimensions. This completeness proof is actually rather simple.
In a unified YMESGT, the vector fields $A_{\mu}^{\ti}$ (and with them
the embedding coordinates $h^{\ti}$) transform, by definition,
 in the adjoint representation of a
simple group $K$. The tensor $C_{\ti\tj\tk}$ of the supergravity theory
then has to be a symmetric cubic
invariant of the adjoint representation of $ K$. Of all the simple groups,
only the unitary groups $SU(N)$ $(N\geq3)$ and
their  different real forms  have such an invariant, namely the Gell-Mann
$d$-symbols
$d_{\ti\tj\tk}=1/2 \textrm{Tr}(T_{\ti}\{ T_{\tj},T_{\tk}\} )$, where $T_{\ti}$
denote the generators of $SU(N)$ or
one of its different real forms.
Just as we did in Section 4, one can then show that only the groups of the
type $SU(N,1)$ ($N \geq 2$) can lead
to positive metrics $g_{\tx\ty}$ and ${\stackrel{\circ}{a}}_{\ti\tj}$,
because only their $d$-symbols can be
transformed to the canonical basis (\ref{canbasis}). Comparing with the
$d$-symbols (\ref{dsymbols}) for the
Jordan algebra $J_{(1,N)}^{\mathbb{C}}$, and taking into account the
isomorphism between its traceless elements
and the generators of $SU(1,N)$,  we see   that the unified YMESGTs we
have just found arise from the gauging of
the unified MESGTs related to $J_{(1,N)}^{\mathbb{C}}$. As the magical
MESGT corresponding to the Euclidean
algebra $J_{3}^{\mathbb{H}}$ is equivalent to the one obtained from
$J_{(1,3)}^{\mathbb{C}}$ (cf the discussion at the end of Section 4 and
Section 6), this shows that
\emph{all} unified  $\CN=2$   MESGTs in five dimensions are obtained by
gauging the full $SU(N,1)$ automorphism groups
of the unified MESGTs defined by the Jordan algebras
$J_{(1,N)}^{\mathbb{C}}$ (see Table 1). It is easy to
convince oneself that the other novel families of unified MESGTs
cannot lead to unified YMESGTs.

Let us close this subsection with a few remarks on the physical
properties of the unified YMESGTs.
In gauging the full $SU(N,1)$ symmetry, the $F \wedge F \wedge A$
term gets replaced by the Chern-Simons form (\ref{CSForm})
 of $SU(N,1)$.  Since the fifth
homotopy group $\Pi_5$ of $SU(N,1)$ is the set of integers $\mathbb{Z}$:
\[ \Pi_5 (SU(N,1))=\Pi_5 (U(N))=\Pi_5(SU(N)) = \mathbb{Z} \ ,  \]
the quantum gauge invariance under  large gauge
transformations require that the dimensionless
 ratio $\frac{g^3}{\kappa}$  of the third power of the
 non-Abelian gauge
coupling constant $g$ and the gravitational constant $\kappa$
must be quantized \cite{GST4}.

As mentioned earlier, a YMESGT without tensor or hypermultiplets 
does not have a scalar potential. This means
that the vacuum expectation values (vevs) of the scalar fields 
$\varphi^{\tx}$ (or equivalently the vevs of the fields
$h^{\ti}(\varphi)$) are not fixed. A vev $\langle h^{\ti} \rangle$ 
corresponding to a compact direction in the
Lie algebra of $SU(N,1)$, can always be chosen as the base point $c$ 
of the canonical basis. The little group of
the base point corresponding to the element $T_0$ of 
$J_{0(1,N)}^{\mathbb{C}}$ is $U(N)\subset SU(1,N)$. This is
the remaining unbroken gauge group in the vacuum. Under 
this unbroken $U(N)$, the $N(N+2)-1$ scalar fields
 decompose
 as $N^{(+1)} \oplus \bar{N}^{(-1)} \oplus (N^2-1)^{(0)} $,
 while the $N(N+2)$ vector fields decompose
 as $1^{(0)} \oplus N^{(+1)} \oplus \bar{N}^{(-1)} \oplus (N^2-1)^{(0)}$.
The singlet is to be identified with the graviphoton, which is thus
no longer `unified' with the other vector
fields under the action of $U(N)$. This was to be expected, as
the non-compact gauge symmetries, which connect
the
 graviphoton with the other vector fields, have
to be broken in any vacuum, as required by  unitarity 
\footnote{In any given Minkowski vacuum with constant vevs
$\langle h^{\ti} \rangle $, the physical graviphoton, i.e., the 
linear combination of vector fields that appears in the
gravitino supersymmetry variation, is given  
by $A_{\mu} = \langle h_{\ti} \rangle A^{\ti}_{\mu}$, where
$h_{\ti}\equiv {\stackrel{\circ}{a}}_{\ti\tj} h^{\tj}$ \cite{GST1}. It 
is easy to see that this linear combination is
automatically invariant under the transformations (\ref{hAtrafo}). 
What we mean when we say that the `unifying'
symmetry maps the `graviphoton' to the other vector fields is that 
it acts irreducibly on all the vector fields
$A_{\mu}^{\ti}$.}

The gauge fields associated with
the non-compact generators eat the scalar fields in the
$N^{(+1)} \oplus \bar{N}^{(-1)}$ and   become massive vector
 fields transforming
 in the $N^{(+1)} \oplus \bar{N}^{(-1)}$ of
 $U(N)$. Due to the extra Yukawa coupling
term  $ \mathcal{L}'$ (eq. (\ref{Yukawaterm}))
 introduced in the Lagrangian to restore supersymmetry
after the  gauging,  the spin $1/2$
 fields in the $N^{(+1)}\oplus \bar{N}^{(-1)}$ of $U(N)$ also become massive.
Together with the massive vector fields, they  form
 massive BPS vector multiplets.
 The central charge of these BPS multiplets is
 generated by the $U(1)$ factor in $U(N)$, which is gauged by the graviphoton.
 The massless spectrum thus consists
 of $\mathcal{N}=2$ $SU(N)$ super
 Yang-Mills coupled to $\mathcal{N}=2$ supergravity.


\subsection{Coupling of Tensor Fields to Unified  Yang-Mills-Einstein
  Supergravity Theories}
As we have seen in the previous section, the only possible gauge groups
of 5D  unified YMESGTs are of the form $SU(N,1)$ with $N\geq 2$. All
these theories are obtained by gauging the full $SU(N,1)$
automorphism groups of the unified MESGTs defined by the Minkowskian
Jordan algebras $J_{(1,N)}^{\mathbb{C}}$. A natural question to ask
now is whether there are gaugings of the
other novel unified MESGTs based on the Jordan algebras
$J_{(1,N)}^{\mathbb{R}}$, $J_{(1,N)}^{\mathbb{H}}$ and $J_{(1,2)}^{\mathbb{O}}$
that come as close as possible to what we called `unified YMESGTs'.
As we have already stated,
one cannot gauge these theories such that all the vector fields
of the ungauged theory become the gauge fields of a simple gauge group.
As was pointed out in \cite{GZ1}, however, if the   $(\tn+1)$-dimensional
 representation of the global symmetry group $G$ of a MESGT decomposes
under a subgroup $K\subset G$ as
\begin{equation}
\mathbf{(\tn+1)_{G}}\rightarrow \textrm{adjoint}(K)\oplus
\textrm{non-singlets}(K) \ ,
\end{equation}
one can sometimes gauge $K$ by turning the non-singlet vector fields
into self-dual tensor fields of the type first described in
\cite{PTvN}. To this end, one splits the index $\ti$ of the vector
fields into two sets:
   $\ti\rightarrow (I,M)$, where $I,J,\ldots$ correspond to 
the adjoint of $K$,
and $M,N,\ldots$ label the non-singlets outside the adjoint. 
The vector fields $A_{\mu}^{I}$ then play the role
of the gauge fields of $K$, whereas the non-singlet vector 
fields $A_{\mu}^{M}$ have to be converted to 2-form
fields $B_{\mu\nu}^{M}$. The technical details of this kind 
of gauging can be found in \cite{GZ1}. One finds that
the gauging of $K$ is possible only if the non-singlets 
transform in a symplectic representation of the gauge
group $K$, and the $C_{\ti\tj\tk}$ components of the type 
$C_{MIJ}$ and $C_{MNP}$ vanish \cite{GZ1}\footnote{As
was pointed out in \cite{Bergshoeffetal},  there are cases 
where the coefficients $C_{MIJ}$ might be non-zero.
For \emph{simple} (i.e., unifying) gauge groups $K$, however, 
this cannot be the case.}. Furthermore, one finds
that the gauging in the presence of tensor fields introduces a 
scalar potential (as well as Yukawa couplings). 
The scalar potential is manifestly positive 
semidefinite \cite{GZ1}.

One can thus try to gauge some of the other unified MESGTs such that the
gauge group $K$ is of the form
$SU(N,1)$, and all the vector fields outside  the adjoint of $K$ are
converted to tensor fields. Such a theory
can then be interpreted as a unified YMESGT coupled to tensor multiplets.
In a vacuum of such a theory, the gauge
group is again  broken down to its maximal compact subgroup $U(N)$,
and the massless spectrum consists of an
$SU(N)$ super Yang-Mills multiplet plus the 5D, $\CN=2$ supergravity
multiplet. The massive part of the spectrum
will consist of $2N$ massive BPS vector multiplets in the
$(N\oplus \bar{N})$ of $SU(N)$ plus
the tensor multiplets, which form massive BPS tensor
multiplets
 \cite{Strathdee,Proceedings,EGZ}.
The number of these tensor multiplets depends on the particular 
theory under consideration
\footnote{Interestingly, if, in any of the above-mentioned theories, one 
chooses to gauge only the $U(N)$ subgroup of
$SU(1,N)$, the $2N$ massive BPS vector multiplets are replaced by 
$2N$ massive BPS tensor multiplets. For the theory based on 
$J_{(1,N)}^{\mathbb{C}}$ with $N=5$, this kind of gauging would, for example, 
  lead to
a minimal 5D, $\mathcal{N}=2$ supersymmetric
$SU(5)$ GUT model with 5D tensor multiplets in the $(5\oplus \bar{5})$ 
of $SU(5)$. A very  similar model was considered in ref. \cite{DGKL},
where the $(5\oplus \bar{5})$ tensor multiplets were interpreted as
the $SU(5)$ 
multiplets that contain the Standard Model Higgs fields. The essential 
difference between the model of \cite{DGKL} and our construction
 is an additional $SU(5)$ singlet
vector multiplet, which was introduced in \cite{DGKL}, 
but does not occur in our case.
It should also be noted that 
 5D $SU(5)$ models with tensor multiplets in the $(5\oplus \bar{5})$
automatically have an additional $U(1)$ factor in the gauge group under 
which the tensor multiplets are charged (see \cite{EGZ} for a general 
proof of this statement). The corresponding gauge field
can be identified with the graviphoton. 
Note finally
that, for any of the theories of the type described above, one can 
always choose to gauge only a subgroup of the
type $SU(1,M)$ with $M<N$. This would simply result in more tensor fields.}.
 Let us first start with the family
of theories based on the Jordan algebras $J_{(1,N)}^{\mathbb{H}}$. 
Under the automorphism group $USp(2N,2)$ of
$J_{(1,N)}^{\mathbb{H}}$, the traceless elements corresponding to 
the vector fields in the MESGT transform in the
anti-symmetric symplectic traceless representation

\[ J_{0(1,N)}^{\mathbb{H}} \Longleftrightarrow (2N^2+3N)  \ .  \]
Since the Jordan algebra $J_{(1,N)}^{\mathbb{H}}$ contains the complex 
Jordan algebra $J_{(1,N)}^{\mathbb{C}}$ as
a subalgebra, one can gauge an $SU(N,1)$ subgroup of $USp(2N,2)$ with 
the remaining $N(N+1)$ vector fields
dualized to tensor fields transforming in the reducible symplectic 
representation
\[ \frac{N(N+1)}{2} \oplus
\overline{\frac{N(N+1)}{2}} \] of $SU(N,1)$ for $N \geq 2$.

In the family of unified MESGTs based on  $J_{(1,N)}^{\mathbb{R}}$,
 the vector fields transform in the symmetric
tensor representation of the automorphism group $SO(N,1)$ of 
$J_{(1,N)}^{\mathbb{R}}$. Now
$J_{(1,N)}^{\mathbb{R}}$ is a subalgebra of $J_{(1,N)}^{\mathbb{C}}$ 
and its traceless elements correspond to the
generators  belonging to  the coset space $SU(N,1)/SO(N,1)$. 
In this case, the maximal unifying gauge groups of the type
$SU(1,M)$ are smaller than the maximal possible non-Abelian gauge groups,
which turn out to be compact.
More concretely, for $N=2n$  with $N>3$ one
can gauge the $U(n)$ subgroup of $SO(2n,1)$ with the remaining 
vector fields dualized to tensor fields
transforming in the reducible symplectic representation
\[ \frac{n(n+1)}{2} \oplus \overline{\frac{n(n+1)}{2}} \]
of $U(n)$. For $N=2n+1$ ($N>3$) one can gauge the $U(n)$ subgroup of
$SO(2n+1,1)$ with the tensor fields  in the
reducible representation
\[ ( n \oplus \bar{n} ) \bigoplus (\frac{n(n+1)}{2} \oplus
\overline{\frac{n(n+1)}{2}}) \bigoplus ( 1 \oplus
\bar{1}) \ .  \] As the gauge groups are not of the type
$SU(1,M)$, the Yang-Mills sectors of these
maximally gauged   theories are, of course, no longer `unified'.

Finally, the novel exceptional octonionic MESGT based on
$J_{(2,1)}^{\mathbb{O}}$ has 26 vector fields
transforming irreducibly under its automorphism group $F_{4(-20)}$.
In this case, one can gauge the $SU(2,1)$ subgroup with the
remaining vector fields replaced by tensor fields transforming in the
reducible symplectic representation
\[ (3 \oplus \bar{3} ) \bigoplus (3 \oplus \bar{3} ) \bigoplus
(3 \oplus \bar{3} ) \ .  \]


\section{The geometry of the novel unified MESGTs and the remarkable
isoparametric hypersurfaces of Elie
Cartan}


\setcounter{equation}{0} In this section we take a first look at 
the geometry of the novel scalar manifolds we
found in this paper. To begin with, let us first isolate the scalar 
manifolds that are symmetric or homogeneous
spaces. From the known classification of $\mathcal{N} =2$ MESGTs 
whose scalar manifolds are symmetric or
homogeneous spaces we expect the scalar manifolds of the novel 
unified MESGTs given in the Section 4 to be
neither symmetric nor homogeneous, in general. More precisely,  
the symmetries of the  $\mathcal{N} =2$ MESGTs
whose scalar manifolds are homogeneous, but not symmetric,  as 
classified in \cite{dWvP1,dWvP2}, are such that
they cannot coincide with any of the novel unified theories 
listed in section 4, since their vector fields 
do not transform irreducibly under a simple noncompact 
symmetry group. Thus, if there are homogeneous
space examples among the novel theories, they also have to be 
symmetric spaces. We already  showed   in Section 3,
that, among the theories with symmetric target spaces,   only the 
four magical MESGTs are unified MESGTs.
Thus, this only leaves open the possibility that some of the novel
unified MESGTs may be equivalent to some of the magical theories.
The magical MESGTs are all defined by cubic
forms that are norm forms of simple Jordan algebras of degree three, so to
answer this question we need to check
if any of the cubic forms of the novel theories coincide with the
norm forms of some simple Jordan algebra of
degree three. Remarkably, and as was already announced at the end of
 Section 4,  we  find that the cubic forms defined by
the structure constants of the traceless
elements of the Minkowskian Jordan algebras $J_{(1,3)}^{\mathbb{R}},
J_{(1,3)}^{\mathbb{C}}$ and
$J_{(1,3)}^{\mathbb{H}}$ coincide with the norm forms of the simple
Jordan algebras $J_{(3)}^{\mathbb{C}},
J_{(3)}^{\mathbb{H}}$ and $J_{(3)}^{\mathbb{O}}$, respectively.
To see this, consider
the example of $J_{(1,3)}^{\mathbb{R}}$,
a general traceless element of which can be parameterized as
\[
 M=\left(
\begin{array}{cccc}
3a & x & y & z \\
-x & -a+b & u & v \\
-y & u & -a+c & w \\
-z & v & w & -a-b-c%
\end{array}%
\right) \ .  \] On the other hand, consider   a  general element of
$J_{(3)}^{\mathbb{C}}$

\[
J=\left(
\begin{array}{ccc}
2a+b & u+iz & v-iy \\
u-iz & 2a+c & w+ix \\
v+iy & w-ix & 2a-b-c%
\end{array}%
\right) \ .
\]
One finds    that
\[
\textrm{det}J=\frac{1}{3}\textrm{tr}M^{3} \ ,
\]
 proving the equivalence of the theories defined by the 
corresponding
cubic forms. One can similarly show the equivalence of the cubic 
forms over the traceless elements of
$J_{(1,3)}^{\mathbb{C}}$ and $J_{(1,3)}^{\mathbb{H}}$ with the 
norm forms of $J_{(3)}^{\mathbb{H}}$ and
$J_{(3)}^{\mathbb{O}}$, respectively \cite{alfa}.
Thus the MESGTs defined by these cubic forms have the 
reduced
structure groups of the simple Jordan algebras 
$J_{(3)}^{\mathbb{C}}, J_{(3)}^{\mathbb{H}}$ and
$J_{(3)}^{\mathbb{O}}$ as their enlarged hidden symmetry groups. 
More specifically, the MESGT defined by
the structure constants of the traceless elements 
of $J_{(1,3)}^{\mathbb{R}}$ has the symmetry group $SL(3,
\mathbb{C})$ which has the automorphism group $SO(3,1)$ of 
$ J_{(1,3)}^{\mathbb{R}}$ as a subgroup. The MESGT
theory defined by the structure constants of $J_{(1,3)}^{\mathbb{C}}$ 
has the symmetry group $SU^*(6)$ which has
the automorphism group $SU(3,1)$ of $J_{(1,3)}^{\mathbb{C}}$
 as a subgroup. Finally, the theory defined by the structure constants of
the traceless elements of
$J_{(1,3)}^{\mathbb{H}}$ has the symmetry group $E_{6(-26)}$, which has
the automorphism group $USp(6,2)$ of
$J_{(1,3)}^{\mathbb{H}}$ as a subgroup \footnote{ $USp(6,2)$ has
$USp(6)\times USp(2)$ as a maximal compact
subgroup.}. To sum up, only three of our novel theories have homogeneous
scalar manifolds, and they are equivalent to three of the theories found in
\cite{GST1}.

Thus, almost all of the novel theories have non-homogeneous scalar manifolds.
Nevertheless, some of these
 non-homogeneous spaces can be  related  to homogeneous spaces in  an
interesting way that resembles an old construction by E. Cartan \cite{cartan}.
More explicitly, consider   the
scalar manifolds of the novel MESGTs defined by the structure
constants of the  Minkowskian Jordan algebras
$J_{(1,2)}^{\mathbb{A}}$
 of degree
 three. They  turn out to be submanifolds
of  ``Minkowskian'' versions of  the target spaces of the magical
supergravity theories defined by the norm forms
of Euclidean Jordan algebras $J_3^{\mathbb{A}}$. These submanifolds 
are themselves
foliated by hypersurfaces that are the
non-compact analogs of the remarkable families of isoparametric
hypersurfaces in 4, 7, 13 and 25 dimensions that
were studied by Elie Cartan long time ago \cite{cartan}.

To show this connection with the work of Cartan, consider the general
element of the Jordan algebra
$J_{(1,2)}^{\mathbb{A}}$ \footnote{ Our labelling follows that of
Cartan, even though he did not use the language
of Jordan algebras.}

\begin{equation}
J=\left(
\begin{array}{ccc}
\sqrt{3}x_{4}-x_{0}-2\cos t & \sqrt{3}x_{3} & -\sqrt{3}\bar{x}_{2} \\
\sqrt{3}\bar{x}_{3} & -\sqrt{3}x_{4}-x_{0}-2\cos t & -\sqrt{3}x_{1} \\
\sqrt{3}x_{2} & \sqrt{3}\bar{x}_{1} & 2x_{0}-2\cos t
\end{array}
\right) \ ,
\end{equation}
where $x_1, x_2, x_3$ are elements of the division algebra
$\mathbb{A}$ and $x_0, x_4$ and $t$ are some real
numbers. The cubic norm of $J$ is given by
\begin{eqnarray} N(J)&=&-8\cos ^{3}t+6\cos
t[-x_{1}\bar{x}_1-x_{2}\bar{x}_2+x_{3}\bar{x}_3+x_{0}^{2}+x_{4}^{2}]
+2x_{0}^{3}-\\ \nonumber  & &
-3x_{0}[2x_{4}^{2}+2x_{3}\bar{x}_3+x_{1}\bar{x}_1+x_{2}
\bar{x}_2]+3
\sqrt{3}x_{4}[x_{1}\bar{x}_1-x_{2}\bar{x}_2]-6\sqrt{3}\textrm{Re}
(x_{3}x_{2}x_{1}) \ ,
\end{eqnarray}
where $\textrm{Re}(x)$ stands for the real part of an element $x$ of
$\mathbb{A}$. Now the hypersurfaces defined by the
condition
\[ N(J)=1 \]
are the coset spaces:
\[ \frac{SL(3,\mathbb{R})}{SL(2,\mathbb{R})} \Leftrightarrow
J_{(1,2)}^{\mathbb{R}} \]
\[ \frac{SL(3,\mathbb{C})}{SU(2,1)} \Leftrightarrow J_{(1,2)}^{\mathbb{C}}  \]
\[ \frac{SU^*(6)}{USp(4,2)} \Leftrightarrow J_{(1,2)}^{\mathbb{H}} \]
\[ \frac{E_{6(-26)}}{F_{4(-20)}} \Leftrightarrow J_{(1,2)}^{\mathbb{O}} \ . \]
Note that in the corresponding manifolds of the magical MESGTs defined by
Euclidean Jordan algebras of degree
three (eq. (\ref{magicals})), 
the reduced structure group is the same, but the automorphism group
is its maximal \emph{compact} subgroup. If
one were to use the cubic norm form  of the above Minkowskian Jordan algebra of
degree three to construct a MESGT, the
kinetic energy terms of the vector fields as well as those  of the scalar
fields would not be positive definite,
rendering these theories  unphysical. 
For the theories constructed in Section 4, on 
the other hand, 
 the cubic polynomials
associated with the novel MESGTs  are given by 

\[ \textrm{tr} J_0^3 \ ,  \]
where $J_0$ is a generic traceless element of $J_{(1,2)}^{\mathbb{A}}$.
Using the parametrization
\[
J_0=\left(
\begin{array}{ccc}
\sqrt{3}x_{4}-x_{0} & \sqrt{3}x_{3} & -\sqrt{3}\bar{x}_{2} \\
\sqrt{3}\bar{x}_{3} & -\sqrt{3}x_{4}-x_{0} & -\sqrt{3}x_{1} \\
\sqrt{3}x_{2} & \sqrt{3}\bar{x}_{1} & 2x_{0}%
\end{array}%
\right) \ ,
\]
one obtains
\begin{eqnarray}
\frac{1}{3} \textrm{tr}
(J_{0}^{3})&=&2x_{0}^{3}-3x_{0}(2x_{4}^{2}+2x_{3}\bar{x}_3+x_{1}
\bar{x}_1+x_{2}\bar{x}_2)+3\sqrt{3}
x_{4}(x_{1}\bar{x}_1-x_{2}\bar{x}_2) -6\allowbreak \sqrt{3}
\textrm{Re} (x_{3}x_{2}x_{1})\nonumber  \\
 & =&\textrm{det}(J)+8\cos ^{3}t-6\cos
t[-x_{1}\bar{x}_1-x_{2}\bar{x}_2+x_{3}\bar{x}_3+x_{0}^{2}+x_{4}^{2}] \ .
\end{eqnarray}
If we now  impose the constraint:
\begin{equation}
-x_{1}\bar{x}_1-x_{2}\bar{x}_2+x_{3}\bar{x}_3+x_{0}^{2}+x_{4}^{2} = 
\frac{4}{3} \cos^2t
\end{equation}
 we find that
\[\frac{1}{3} \textrm{tr}(J_{0}^{3}) = \det(J)|_{\rm constraint(6.4)} \]
Noting that for a fixed value of $t$ the equation $(6.4)$ 
defines a non-compact ``hypersphere'' and comparing our
formulas with those of Elie Cartan, we see that the equation
\[ \det J =\textrm{constant} \]
subject to the constraint $-x_{1}\bar{x}_1-x_{2}\bar{x}_2+x_{3}\bar{x}_3+
x_{0}^{2}+x_{4}^{2} = \frac{4}{3}
\cos^2t $  go over to his equations if we replace $x_1\bar{x}_1$ and 
$x_2\bar{x}_2$ with their negatives. Thus,
the scalar manifolds of the novel MESGTs defined by 
$J_{(1,2)}^{\mathbb{A}}$ are the noncompact analogs of the
manifolds studied by Cartan.  Cartan showed that the remarkable 
compact hypersurfaces he studied in 4, 7, 13 and
25 dimensions  exhaust the list of isoparametric hypersurfaces in 
spheres with three distinct curvatures.
 Since the scalar manifolds of the novel
unified theories are given  by the condition
\[\textrm{tr}(J_{0}^{3})=
\textrm{constant} \ ,  \] we see that they are foliated  by the noncompact 
analogues of these remarkable
hypersurfaces.

The hypersurfaces studied by Cartan are related to the homogeneous spaces 
\cite{wuwu}
\begin{eqnarray}
\frac{SO(3)}{\mathbb{Z}_2^2} \\ \nonumber \frac{SU(3)}{T^2} \\ \nonumber 
\frac{USp(6)}{SU(2)^3} \\
\nonumber \frac{F_4}{Spin(8)}
\end{eqnarray}
 The corresponding  hypersurfaces in the scalar manifolds 
of MESGTs defined by $J_{(1,2)}^{\mathbb{A}}$  are
related to certain noncompact versions of the above homogeneous 
manifolds. The detailed study of these manifolds
and their extension to manifolds of theories defined by higher 
dimensional Jordan algebras will be left for future work.


\section{Conclusions}

Extended supergravity theories often exhibit  extra non-compact 
bosonic symmetries that are not part of
the underlying R-symmetry groups. These extra symmetries can 
connect fields with the same Lorentz
quantum numbers even if these fields originate from different types of 
supermultiplets. In particular, there can
exist non-compact bosonic symmetries that can mediate  between vector 
fields from vector multiplet sectors and
vector fields from the supergravity sector. In some cases, all the 
vector fields transform irreducibly under a
single simple symmetry group of this type. Alluding to the conventional 
GUT terminology, we called such
supergravity theories `unified' MESGTs or `unified' YMESGTs depending on 
whether the simple, `unifying' symmetry
group is a global or a local symmetry of the theory. For 5D,   
$\mathcal{N}=4$ supergravity, such a unifying
symmetry is impossible, as there would  always be at least one singlet 
vector field in the $\mathcal{N}=4$
supergravity multiplet, no matter how the simple symmetry group is 
chosen \cite{5DN4}. Remarkably, for  5D,   $\mathcal{N}=2$
supergravity, such unified MESGTs and YMESGTs do exist.

 The general 5D,   $\mathcal{N}=2$ MESGTs are in
one-to-one correspondence with cubic polynomials 
$\mathcal{V}(h)=C_{\ti\tj\tk}h^{\ti}h^{\tj}h^{\tk}$ that can be
brought to the canonical form (\ref{canbasis}), 
which ensures positive kinetic terms in the action. A unified
MESGT is obtained, when such an admissible set of coefficients
  $C_{\ti\tj\tk}$ forms an invariant
symmetric tensor of an irreducible representation of a simple group.

In this paper, we have found infinitely many examples of such tensors by using
the language of Jordan algebras. There are essentially two ways a Jordan
algebra can give rise to a cubic polynomial of the type
$\mathcal{V}(h)=C_{\ti\tj\tk}h^{\ti}h^{\tj}h^{\tk}$:\\
(i) Every Jordan algebra $J$ has a norm form $N:J\rightarrow \mathbb{R}$.
When this
norm form is cubic (i.e., when the Jordan algebra is of degree $p=3$), it
defines
a cubic polynomial $\mathcal{V}(h)=N$. In general, such a cubic polynomial
can not be brought to the canonical form (\ref{canbasis}). The Jordan algebras
for which this \emph{is} possible, are precisely the \emph{Euclidean}
Jordan algebras of degree three  (cf. items (i) and (ii) in Section 3 or 
Section 4). All these Euclidean Jordan algebras of
degree three
define admissible MESGTs, however, only the four \emph{simple} ones (i.e.
the Hermitian
$(3\times 3)$ matrices over $\mathbb{R}, \mathbb{C}, \mathbb{H}, \mathbb{O}$)
   lead to \emph{unified} MESGTs. These unified MESGTs were constructed
in \cite{GST1}.\\
(ii) Being completely symmetric, the structure constants of a Jordan
algebra also
define a cubic polynomial. By construction, this polynomial is invariant
under the
automorphism group of the Jordan algebra, but as this automorphism group
does not act
irreducibly on the Jordan algebra generators, such a polynomial cannot
give rise to a
\emph{unified} MESGT. In this paper, we showed, that one can nevertheless
obtain
unified MESGTs from these structure constants, provided that one restricts
oneself
to the traceless elements and takes $J$ to be equal to any of the
\emph{Minkowskian}
Jordan algebras $J_{(1,N)}^{\mathbb{A}=\mathbb{R},\mathbb{C},\mathbb{H}}$
or $J_{(1,2)}^{\mathbb{O}}$. This way, we obtained three infinite families
and one
novel exceptional unified MESGT. Interestingly, the novel theories based on
$J_{(1,3)}^{\mathbb{R}}$,  $J_{(1,3)}^{\mathbb{C}}$ $J_{(1,3)}^{\mathbb{H}}$
are equivalent to the ones 
based on $J_{3}^{\mathbb{C}}$, $J_{3}^{\mathbb{H}}$,
$J_{3}^{\mathbb{O}}$, respectively.

The unified MESGTs based on $J_{(1,N)}^{\mathbb{C}}$ can all be
turned into unified YMESGTs by gauging the full automorphism groups $SU(1,N)$.
These theories exhaust all possible
unified YMESGTs in 5D and include the one discovered in \cite{GST2}.

As a by-product of our considerations, we found that the scalar 
manifolds based on $J_{(1,2)}^{\mathbb{A}}$ are
foliated by certain noncompact analogues of the isoparametric 
hypersurfaces in spaces of constant curvature
studied by E. Cartan \cite{cartan} long time ago.

The existence of unified $\mathcal{N}=2$ MESGTs and YMESGTs is not special 
only to five dimensions. They also exist in
four dimensions. The classification of the four dimensional unified 
MESGTs and YMESGTs,  as well as
 the higher-dimensional origin of our theories will be left for future work.

\textbf{Acknowledgements:} One of us (M.G.) would like 
to thank the hospitality of the CERN theory division where
part of this work was done.

\end{document}